\documentclass[USenglish,twocolumn]{article}

\ifx\directlua\undefined\ifx\XeTeXcharclass\undefined
  \usepackage[utf8]{inputenc}                           
  \else\RequirePackage[no-math]{fontspec}[2017/03/31]\fi 
  \else\RequirePackage[no-math]{fontspec}[2017/03/31]\fi 
\usepackage[sort&compress,square,numbers]{natbib}
\usepackage[big,online]{dgruyter}
\usepackage{ragged2e}
\theoremstyle{dgthm}

\theoremstyle{dgdef}

\begin{document}

\articletype{Research Article}

\author[1,2]{Xiaomin Lv†}
\author[1]{Ze Wang†}
\author[1]{Tianyu Xu†}
\author[1]{Chen Yang†*}                 
\author[1]{Xing Jin}
\author[1]{Binbin Nie}
\author[1]{Du Qian}
\author[1]{Yanwu Liu}
\author[1]{Kaixuan Zhu}
\author[1]{Bo Ni}
\author[1,2,3,4]{Qihuang Gong}
\author[5]{Fang Bo*}                   
\author[1,2,3,4]{Qi-Fan Yang*}         

\affil[1]{State Key Laboratory for Artificial Microstructure and Mesoscopic Physics and Frontiers Science Center for Nano-optoelectronics, School of Physics, Peking University, Beijing 100871, China}
\affil[2]{Hefei National Laboratory, Hefei 230088, China}

\affil[3]{Collaborative Innovation Center of Extreme Optics, Shanxi University, Taiyuan 030006, China}
\affil[4]{Peking University Yangtze Delta Institute of Optoelectronics, Nantong 226010, China}
\affil[5]{Nankai University, Tianjin 300071, China}

\title{\textbf{Electrically-pumped soliton microcombs on thin-film lithium niobate}}
\runningtitle{Electrically-pumped soliton microcombs on thin-film lithium niobate}

\abstract{
Thin-film lithium niobate (TFLN) has enabled efficient on-chip electro-optic modulation and frequency conversion for information processing and precision measurement. Extending these capabilities with optical frequency combs unlocks massively parallel operations and coherent optical-to-microwave transduction, which are achievable in TFLN microresonators via Kerr microcombs. However, fully-integrated Kerr microcombs directly driven by semiconductor lasers remain elusive, which has delayed integration of these technologies. Here we demonstrate electrically pumped TFLN Kerr microcombs without optical amplification. With optimized laser-to-chip coupling and optical quality factors, we generate soliton microcombs at a 200~GHz repetition frequency with an optical span of 180~nm using only 25~mW of pump power. Moreover, self-injection locking enables turnkey initiation and substantially narrows the laser linewidth. Our work provides integrated comb sources for TFLN-based communicational, computational, and metrological applications.
}
\journalname{Nanophotonics}
\journalyear{2025}
\journalvolume{aop}

\maketitle
\markboth{Xiaomin Lv et al.: Electrically-pumped soliton microcombs on thin-film lithium niobate}{Xiaomin Lv et al.: Electrically-pumped soliton microcombs on thin-film lithium niobate}
\vspace*{-6pt}

\section{Introduction} 
\vspace*{-3pt}

Lithium niobate (LN) is renowned for its strong second-order nonlinearities, enabling electro-optic modulation, second-harmonic generation, and spontaneous parametric down-conversion. The advent of thin-film lithium niobate (TFLN) has translated these capabilities to photonic chips by supporting low-loss LN integrated circuits \cite{zhu2021integrated,boes2023lithium}. State-of-the-art devices now include high-speed, low-drive-voltage electro-optic (EO) modulators \cite{wang2018integrated, RN3163} and high-efficiency periodically poled LN (PPLN) waveguides that extend toward the ultraviolet \cite{wang2018ultrahigh, Hwang:23}. These components are rapidly transitioning to commercial use across many fields. 

Optical frequency combs (OFCs) comprise evenly spaced spectral lines that, under Fourier transform, correspond to periodic pulse trains \cite{diddams2020optical}. Beyond tabletop mode-locked lasers, chip-scale realizations have emerged in high-$Q$ microresonators, where resonant field enhancement drives nonlinear frequency generation \cite{yang2024efficient}. Two principal mechanisms are employed: electro-optic modulation and Kerr nonlinearity. In electro-optic combs, the line spacing is set by the microwave drive (typically $<100~\mathrm{GHz}$), whereas Kerr microcombs offer broader flexibility with repetition frequency determined by the resonator free-spectral range (FSR) and, owing to relaxed phase-matching constraints, can access wider optical bandwidths. These attributes position Kerr microcombs for applications in parallel communications \cite{marin2017microresonator,jorgensen2022petabit}, optical computing \cite{feldmann2021parallel,xu202111,bai2023microcomb}, and optical clocks \cite{newman2019architecture}. 

On TFLN, Kerr microcombs have typically required amplified continuous-wave pumps to reach the parametric oscillation threshold, hindering full integration \cite{he2019self,lv2025broadband,nie2025soliton,song2024octave}. By contrast, in centrosymmetric, ultralow-loss platforms such as $\mathrm{Si_3N_4}$, direct semiconductor-laser pumped microcombs have been demonstrated and deployed \cite{shen2020integrated,wang2025reconfigurable}. Realizing fully integrated TFLN Kerr microcombs has thus hinged on mitigating losses -- specifically, improving laser-to-chip coupling and achieving sufficiently high $Q$ to enable parametric oscillation at modest pump powers.

\begin{figure*}[htbp!]
    \centering
    \includegraphics[width=\linewidth]{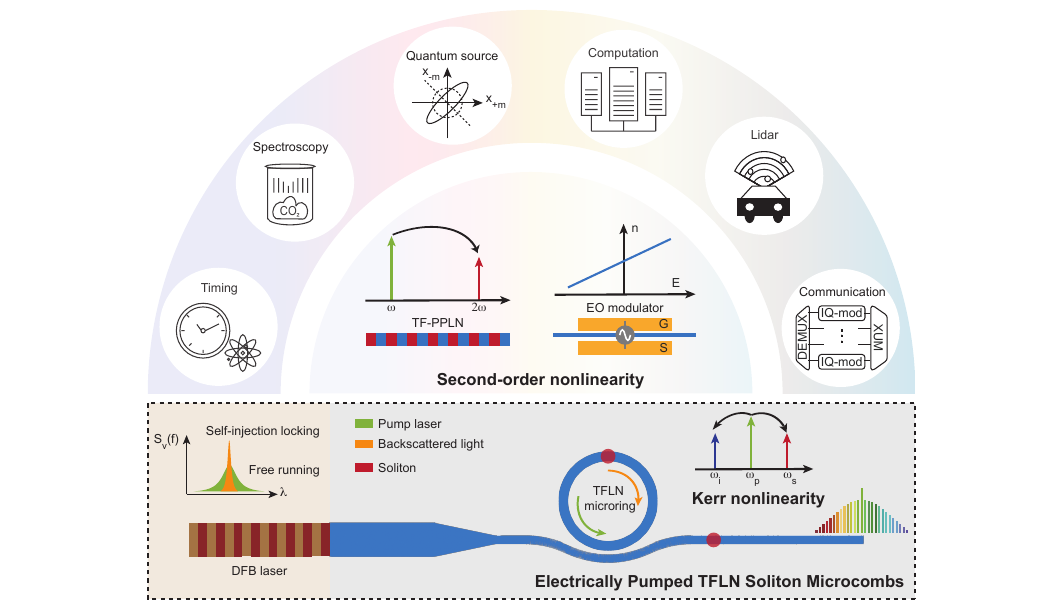 }
    \caption{\justifying\textbf{Electrically pumped TFLN soliton microcombs and representative applications.} }
    \label{figure1}
\end{figure*}

Here, we demonstrate integrated TFLN Kerr microcombs directly driven by a distributed-feedback (DFB) laser (Fig.~\ref{figure1}). Engineered edge couplers deliver approximately $25~\mathrm{mW}$ of on-chip pump power, sufficient to initiate parametric oscillation in a microresonator with $Q>3\times10^{6}$. The operation is robust owing to self-injection locking (SIL), which provides turnkey initiation and narrows the laser linewidth to approximately $4.7~\mathrm{kHz}$. The resulting soliton microcombs feature a $200~\mathrm{GHz}$ repetition frequency and an optical span exceeding $180~\mathrm{nm}$.

\vspace*{-13pt}

\section{Methods}

\subsection{Device design}

\begin{figure}[t]
    \centering
    \includegraphics[width=\linewidth]{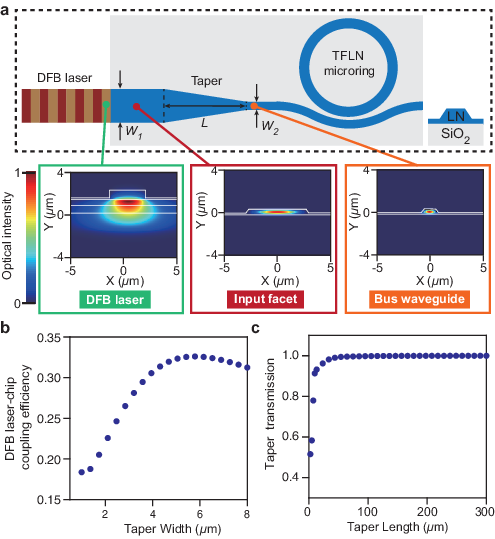}
    \caption{\justifying\textbf{Device layout.}
    (a) The DFB laser and the microresonator. The LN waveguide width is reduced from $W_{1}$ to $W_{2}$ over a length $L$ to enable efficient coupling to the laser. Insets: simulated fundamental transverse-electric (TE) modes of the DFB laser (left), the input facet (middle), and the bus waveguide (right).
    (b) Simulated laser-chip coupling efficiency as a function of the facet size.
    (c) Simulated transition efficiency as a function of the taper length.}
    \label{figure2}
\end{figure}

The detailed device configuration is shown in Fig.~\ref{figure2}a. The DFB laser (DC-1550-HP-02 V008, Shijia) emits a near-Gaussian mode with a large mode-field diameter owing to weak index confinement. The microresonator is fabricated on a z-cut LN on-insulator wafer with a LN layer thickness of 600~nm, partially etched by 390~nm, with a wedge angle of approximately 60°, forming a ridge width of 1.2~\,$\mu\mathrm{m}$. To mitigate photorefraction, the device is air-clad \cite{xu2021mitigating}. Because the mode of the bus waveguide (1~\,$\mu\mathrm{m}$ wide) is severely mismatched to the laser mode, a taper is introduced: the waveguide width is expanded at the facet and then transitions adiabatically to the narrow bus waveguide. Numerical simulations predict an optimal laser-to-chip coupling efficiency of $33\%$ at $W_{1}=5.8~\mu\mathrm{m}$ (Fig.~\ref{figure2}b). To suppress multimode excitation while maintaining high efficiency, we select $W_{1}=3.3~\mu\mathrm{m}$, which yields a weaker higher-order mode content and an acceptable coupling efficiency of $28\%$. Adiabatic behavior is achieved to set the transition length $L\gtrsim100~\mu\mathrm{m}$ (Fig.~\ref{figure2}c); in practice we choose $L=\,$300~\,$\mu\mathrm{m}$.

Given the DFB laser output power of $\approx170~\mathrm{mW}$ at 500~mA, the maximum available on-chip pump power is $<50~\mathrm{mW}$. This budget imposes a stringent requirement on the parametric oscillation threshold $P_\mathrm{th}$: empirically, comb initiation is reliable when the pump exceeds $(3$–$4)\times P_\mathrm{th}$. We therefore set $P_\mathrm{th}\le 12.5~\mathrm{mW}$ as a design target.

\begin{figure}[t]
    \centering
    \includegraphics[width=\linewidth]{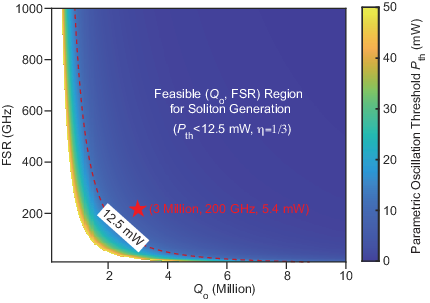}
    \caption{\justifying\textbf{Design optimization for DFB-laser–pumped parametric oscillation.}
    Threshold power map $P_\mathrm{th}$ as a function of intrinsic quality factor $Q_o$ and FSR under the assumption of optimal coupling $\eta=1/3$ . The blue region indicates the soliton generation space, bounded by the red dashed line at $P_\mathrm{th}=12.5~\mathrm{mW}$. The star marker denotes the target operating point ($Q_o=3\times10^{6}$, $\mathrm{FSR}=200~\mathrm{GHz}$) with a calculated threshold of $5.4~\mathrm{mW}$.}
    \label{figure3}
\end{figure}

The parametric oscillation threshold is given by \cite{kippenberg2004kerr}
\begin{equation}
P_\mathrm{th}
= \frac{\pi n \,\omega_0 A_\mathrm{eff}}{4n_2}\frac{1}{\eta(1-\eta)^2}\frac{1}{D_1 Q_\mathrm{o}^2},
\end{equation}
where $Q_o$ is the intrinsic quality factor, $n=2.21$ is the TFLN refractive index, $\omega_0/2\pi=193$ THz is the pump frequency, $A_\mathrm{eff}=0.765~\mu\mathrm{m}^2$ is the effective mode area, $n_2=1.8\times10^{-19}~\mathrm{m}^2/\mathrm{W}$ is the Kerr nonlinear index, and $D_1/2\pi$ is the FSR. The loading factor $\eta$ describes the portion of coupling losses in total loss; to minimize $P_\mathrm{th}$, we found $\eta=1/3$ is the optimal condition.

Figure~\ref{figure3} presents the calculated threshold for the TFLN microresonator as a function of $Q_o$ and FSR under optimal coupling. Increasing $Q_o$ significantly reduces $P_\mathrm{th}$ and broadens the accessible range of FSRs for a fixed pump budget, whereas lower $Q_o$ necessitates larger FSRs to maintain low thresholds. For our typical fabrication performance ($Q_o\approx3\times10^{6}$; see Sec.~3.1), the analysis indicates that $\mathrm{FSR}\gtrsim200~\mathrm{GHz}$ enables robust soliton generation within the $50~\mathrm{mW}$ budget. We thus chose 200-GHz FSR for the microresonator.

\subsection{Device Fabrication}
We employed an optimized fabrication process to realize high-Q resonators on commercial Z-cut TFLN on insulator chips (NANOLN)\cite{RN3784}.  The devices were patterned by electron beam lithography using an $800\,\mathrm{nm}$-thick hydrogen silsesquioxane (HSQ) resist. A multipass writing technique was used to ensure high-quality pattern definition. The patterns were subsequently etched by argon-ion (Ar$^{+}$)-based inductively coupled plasma (ICP). After etching, the residual HSQ was removed using diluted hydrofluoric acid. Deposited residues were eliminated by performing two consecutive cleanings in Standard Clean~1 (SC-1) solution (a mixture of NH$_4$OH, H$_2$O$_2$, and H$_2$O) heated to $\sim 85\,^{\circ}\mathrm{C}$. Finally, bus waveguide facets were exposed by manual cleaving.

\section{Results}

\subsection{Device Characterization}

\begin{figure}[htbp!]
    \centering
    \includegraphics[width=\linewidth]{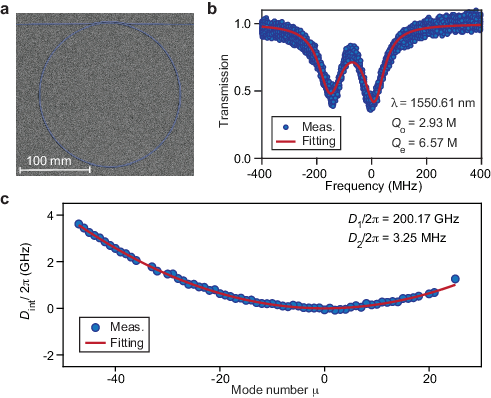}
    \caption{\justifying\textbf{Device characterization.}
    (a) Scanning electron micrograph of the microresonator.
    (b) Normalized transmission near $\lambda=1550.61~\mathrm{nm}$ with a fit yielding $Q_o=2.93\times10^6$ and $Q_e=6.57\times10^6$ (blue: data; red: fit).
    (c) Integrated dispersion $D_\mathrm{int}/2\pi$ of the fundamental TE mode.}
    \label{figure4}
\end{figure}

To characterize the fabricated resonator (Fig.~\ref{figure4}a), we record the transmission spectra using a tunable external-cavity diode laser whose frequency sweep is calibrated by an unbalanced Mach–Zehnder interferometer. The resonance doublet observed in Fig.~\ref{figure4}b arises from Rayleigh backscattering that couples counter-propagating modes \cite{RN4288}. Fitting the doublet lineshape yields an intrinsic quality factor $Q_o=2.93\times10^6$ and a coupling quality factor $Q_e=6.57\times10^6$. From these, the loading factor is $\eta\approx0.3$, corresponding to an estimated parametric oscillation threshold of $P_\mathrm{th}= 5.4~\mathrm{mW}$ for the parameters used in our design model.

We next expand the laser scan to extract the dispersion profile of the fundamental TE family over 1510–1630~nm. The integrated dispersion is defined as
$
D_\mathrm{int}=\omega_\mu-\omega_0-D_1\mu=\sum_{n\ge2}\tfrac{D_n\mu^n}{n!},
$
where $\omega_\mu$ is the resonant frequency of the $\mu$th mode relative to the pumped mode. As shown in Fig.~\ref{figure4}c, we obtain $\mathrm{FSR}=D_1/2\pi=200.17~\mathrm{GHz}$ and, from a polynomial fit to $D_\mathrm{int}$, a second-order dispersion $D_2/2\pi=3.25~\mathrm{MHz}$. This indicates anomalous dispersion that is essential for achieving mode-locking in the form of bright solitons in the microresonator.

To quantify the insertion loss at the laser-chip interface, we first measure the lensed-fiber–to–chip-facet coupling via a fiber–chip–fiber transmission experiment with the laser detuned from cavity resonances. Comparing the measured fiber output with the inferred on-chip power yields a laser-to-chip coupling efficiency of $\sim20\%$ (insertion loss $\approx7~\mathrm{dB}$), modestly lower than the simulated $28\%$ ($\approx5.5~\mathrm{dB}$). We attribute the discrepancy to fabrication tolerances, including facet roughness and slight sidewall tilt introduced during dicing and cleaving.

\begin{figure}[t]
\centering
\includegraphics[width=\linewidth]{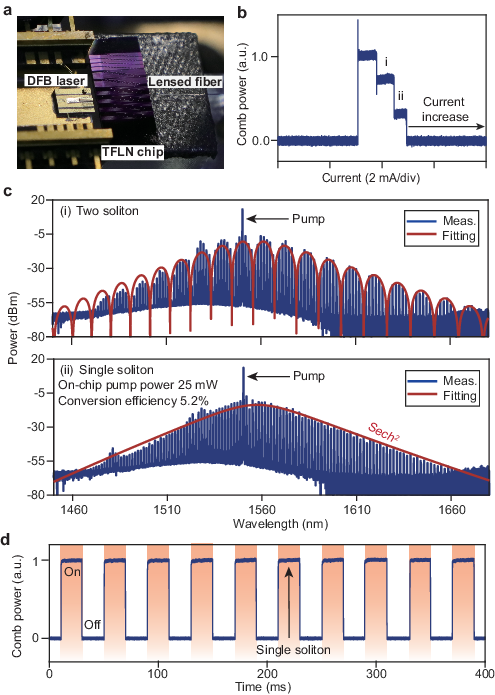}
\caption{\justifying\textbf{Experimental soliton microcomb generation.}
(a) Experimental setup.
(b) Measured comb power versus driving current of the DFB laser.
(c) Optical spectra for (i) a two-soliton state and (ii) a single-soliton state, corresponding to the plateaus in panel (b).
(d) Repeatable turnkey initiation tests.}
\label{figure5}
\end{figure}

\begin{figure*}[t]
\centering
\includegraphics[width=\linewidth]{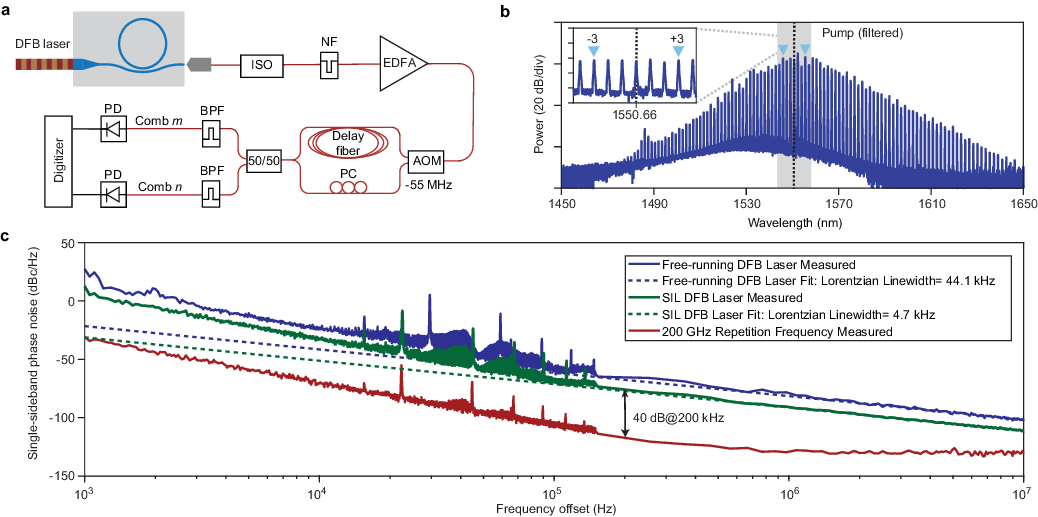}
\caption{\justifying\textbf{Coherence characterization of soliton microcombs.}
(a) Measurement setup. ISO: isolator; NF: notch filter; EDFA: erbium-doped fiber amplifier; AOM: acousto-optic modulator; PC: polarization controller; BPF: band-pass filter; PD: photodetector.
(b) Soliton spectrum after the NF. The $\pm3$ comb lines used for repetition-rate phase-noise measurement are indicated by triangular markers; the filtered pump is shown by the vertical dashed line.
(c) Single-sideband phase noise of the free-running DFB laser, the SIL DFB laser, and the comb repetition frequency.}
\label{Fig3}
\end{figure*}

\subsection{Comb generation}
The photograph of the experimental setup is shown in Fig.~\ref{figure5}a. The DFB laser and the TFLN chip are mounted on independent translation stages, and the laser output is precisely aligned to the bus waveguide; the chip output is collected with a lensed fiber. SIL requires an appropriate optical feedback phase \cite{shen2020integrated}, which we tune by translating the DFB laser on a piezoelectric stage. With proper phase, scanning the DFB laser frequency across a cavity mode produces characteristic step-like features in the comb power (measured after pump suppression by a notch filter), as shown in Fig.~\ref{figure5}b. Three low-noise plateaus are observed reproducibly. Setting the drive current within a given plateau enables deterministic access to the corresponding soliton state.

Representative spectra are shown in Fig.~\ref{figure5}c. The second-highest plateau corresponds to a two-soliton state, which exhibits spectral modulation from soliton interference, while the lowest plateau yields a single-soliton state with a smooth $\mathrm{sech}^2$ envelope. The comb spans $1480$–$1660~\mathrm{nm}$, with central lines reaching -7.4~dBm. Excluding the residual pump, the single-soliton pump-to-comb conversion efficiency is $5.2\%$.

A key benefit of SIL is turnkey initiation of the comb \cite{shen2020integrated}: once the current is preset, solitons form automatically upon laser activation. We emulate repeated start-up by square-wave modulating the drive current (Fig.~\ref{figure5}d). Each cycle deterministically returns the system to the single-soliton state at the chosen current, certifying the robustness of the turnkey initiation.

\subsection{Coherence characterization}
We assess comb coherence using a delayed self-heterodyne (DSH) interferometer (Fig.~\ref{Fig3}a). The setup employs a $1$~km fiber delay and an AOM driven at $55$~MHz by a low-noise source. The undiffracted (zero-order) and first-order AOM outputs form the two interferometer arms, which are recombined and detected. The interference signal is recorded on an oscilloscope; instantaneous frequency fluctuations are extracted via a Hilbert-transform demodulation to obtain the single-sideband phase-noise power spectral density. To measure the phase noise of the soliton repetition frequency, we simultaneously acquire the DSH signals of the $\pm3$ comb lines (Fig.~\ref{Fig3}b) and, after demodulation, compute their differential phase. Using the comb constraint, this directly yields the phase evolution of the repetition frequency \cite{jeong2020ultralow,lao2023quantum}.

To mitigate etalon ripples from the delay fiber at high offsets ($>200$~kHz), the noise is sampled at discrete frequencies $f_n=(2n-1)/(2\tau)$ with $\tau=4.8821~\mu\mathrm{s}$. Figure~\ref{Fig3}c compares the phase noise of the DFB laser in free-running and SIL states, together with the $200$~GHz comb repetition frequency. The free-running laser exhibits a Lorentzian linewidth of $\sim44.1$~kHz, which narrows to $\sim4.7$~kHz under SIL. The comb repetition frequency reaches $-130~\mathrm{dBc/Hz}$ at a $1$~MHz offset. Because at sub-kilohertz offset frequencies the system is sensitive to environmental perturbations, we focus on the $1$~kHz–$10$~MHz range to investigate the noise origin. Below $\sim200$~kHz, the phase-noise spectra of the SIL laser and the repetition frequency display similar features, indicating direct noise transduction from pump to comb.  A quantitative transduction factor is obtained at a $200$~kHz offset, yielding $\sim40$~dB from the SIL-laser phase noise to the repetition frequency phase noise.

\section{Discussion}
Future packaging of the demonstrated integrated soliton microcombs will enable robust operation outside laboratory environments \cite{shen2020integrated, wang2025compact}. Nevertheless, the full potential emerges when co-integrated with other LN components on the same die. For instance, colocating modulators and comb sources supports massively parallel wavelength-division-multiplexed links from chip-to-chip interconnects to long-haul systems, with the latter benefiting from low-noise amplification in Er-doped TFLN waveguides \cite{RN4284}. Besides, completing comb functionality by adding PPLN waveguides enables $f$–$2f$ self-referencing \cite{diddams2020optical}.

Progress toward this vision requires: (i) migration to X-cut LN to maximize electro-optic and second-harmonic efficiencies \cite{nie2025soliton,RN4283}; (ii) higher comb line power, e.g., via dark-pulse operation in normal-dispersion microresonators \cite{xue2015mode,lv2025broadband}; and (iii) broader span through dispersion engineering \cite{song2024octave} and coupled-resonator structures \cite{zhu2025ultra}, together with low-loss claddings that provide thermal tuning while preserving high optical $Q$. With continued advances in microfabrication and a deepening understanding of microcomb dynamics \cite{yang2024efficient}, fully integrated TFLN comb circuits for communications \cite{shu2022microcomb,corcoran2025optical}, portable optical clocks \cite{newman2019architecture}, and artificial intelligence \cite{RN4281} are within reach.

\begin{acknowledgement}
The authors thank Xinrui Luo, Hangzhe Lv, and Zhenyu Xie for helpful discussions.
\end{acknowledgement}

\begin{funding}
This work was supported by National Key R\&D Plan of China (Grant No. 2021ZD0301500) and National Natural Science Foundation of China (12293050, 12304412).
\end{funding}

\begin{authorcontributions}
Xiaomin Lv, Ze Wang, and Tianyu Xu performed the experiments with assistance from Xing Jin and Kaixuan Zhu. Xiaomin Lv designed the microresonator and edge couplers with assistance from Yanwu Liu and Tianyu Xu. Chen Yang fabricated the chip, and Duqian captured the chip photographs. All authors contributed equally to writing the manuscript.
\end{authorcontributions}

\begin{conflictofinterest}
The authors declare no competing interests.
\end{conflictofinterest}

\begin{dataavailabilitystatement}
The datasets generated and analysed during the current study are available from the corresponding author upon reasonable request.
\end{dataavailabilitystatement}

\vspace*{-14pt}

\bibliographystyle{unsrt}  
\bibliography{main}

\begin{thebibliography}{10}

\bibitem{zhu2021integrated}
Di~Zhu, Linbo Shao, Mengjie Yu, Rebecca Cheng, Boris Desiatov, C.J Xin, Yaowen Hu, Jeffrey Holzgrafe, Soumya Ghosh, Amirhassan Shams-Ansari, et~al.
\newblock Integrated photonics on thin-film lithium niobate.
\newblock {\em Adv. Opt. Photonics}, 13(2):242--352, 2021.

\bibitem{boes2023lithium}
Andreas Boes, Lin Chang, Carsten Langrock, Mengjie Yu, Mian Zhang, Qiang Lin, Marko Lon{\v{c}}ar, Martin Fejer, John Bowers, and Arnan Mitchell.
\newblock Lithium niobate photonics: Unlocking the electromagnetic spectrum.
\newblock {\em Science}, 379(6627):eabj4396, 2023.

\bibitem{wang2018integrated}
Cheng Wang, Mian Zhang, Xi~Chen, Maxime Bertrand, Amirhassan Shams-Ansari, Sethumadhavan Chandrasekhar, Peter Winzer, and Marko Lon{\v{c}}ar.
\newblock Integrated lithium niobate electro-optic modulators operating at cmos-compatible voltages.
\newblock {\em Nature}, 562(7725):101--104, 2018.

\bibitem{RN3163}
Mian Zhang, Cheng Wang, Prashanta Kharel, Di~Zhu, and Marko Lončar.
\newblock Integrated lithium niobate electro-optic modulators: when performance meets scalability.
\newblock {\em Optica}, 8(5):652--667, 2021.

\bibitem{wang2018ultrahigh}
Cheng Wang, Carsten Langrock, Alireza Marandi, Marc Jankowski, Mian Zhang, Boris Desiatov, Martin~M Fejer, and Marko Lon{\v{c}}ar.
\newblock Ultrahigh-efficiency wavelength conversion in nanophotonic periodically poled lithium niobate waveguides.
\newblock {\em Optica}, 5(11):1438--1441, 2018.

\bibitem{Hwang:23}
Emily Hwang, Nathan Harper, Ryoto Sekine, Luis Ledezma, Alireza Marandi, and Scott Cushing.
\newblock Tunable and efficient ultraviolet generation with periodically poled lithium niobate.
\newblock {\em Opt. Lett.}, 48(15):3917--3920, Aug 2023.

\bibitem{diddams2020optical}
Scott~A Diddams, {Kerr}y Vahala, and Thomas Udem.
\newblock Optical frequency combs: coherently uniting the electromagnetic spectrum.
\newblock {\em Science}, 369(6501):eaay3676, 2020.

\bibitem{yang2024efficient}
Qi-Fan Yang, Yaowen Hu, Victor Torres-Company, and {Kerr}y Vahala.
\newblock Efficient microresonator frequency combs.
\newblock {\em eLight}, 4(1):18, 2024.

\bibitem{marin2017microresonator}
Pablo Marin-Palomo, Juned~N Kemal, Maxim Karpov, Arne Kordts, Joerg Pfeifle, Martin~HP Pfeiffer, Philipp Trocha, Stefan Wolf, Victor Brasch, Miles~H Anderson, et~al.
\newblock Microresonator-based solitons for massively parallel coherent optical communications.
\newblock {\em Nature}, 546(7657):274--279, 2017.

\bibitem{jorgensen2022petabit}
AA~J{\o}rgensen, D~Kong, MR~Henriksen, F~Klejs, Z~Ye, {\`O}B~Helgason, HE~Hansen, H~Hu, M~Yankov, S~Forchhammer, et~al.
\newblock Petabit-per-second data transmission using a chip-scale microcomb ring resonator source.
\newblock {\em Nat. Photonics}, pages 1--5, 2022.

\bibitem{feldmann2021parallel}
Johannes Feldmann, Nathan Youngblood, Maxim Karpov, Helge Gehring, Xuan Li, Maik Stappers, Manuel Le~Gallo, Xin Fu, Anton Lukashchuk, Arslan~Sajid Raja, et~al.
\newblock Parallel convolutional processing using an integrated photonic tensor core.
\newblock {\em Nature}, 589(7840):52--58, 2021.

\bibitem{xu202111}
Xingyuan Xu, Mengxi Tan, Bill Corcoran, Jiayang Wu, Andreas Boes, Thach~G Nguyen, Sai~T Chu, Brent~E Little, Damien~G Hicks, Roberto Morandotti, et~al.
\newblock 11 {TOPS} photonic convolutional accelerator for optical neural networks.
\newblock {\em Nature}, 589(7840):44--51, 2021.

\bibitem{bai2023microcomb}
Bowen Bai, Qipeng Yang, Haowen Shu, Lin Chang, Fenghe Yang, Bitao Shen, Zihan Tao, Jing Wang, Shaofu Xu, Weiqiang Xie, et~al.
\newblock Microcomb-based integrated photonic processing unit.
\newblock {\em Nat. Commun.}, 14(1):66, 2023.

\bibitem{newman2019architecture}
Zachary~L Newman, Vincent Maurice, Tara Drake, Jordan~R Stone, Travis~C Briles, Daryl~T Spencer, Connor Fredrick, Qing Li, Daron Westly, Bojan~R Ilic, et~al.
\newblock Architecture for the photonic integration of an optical atomic clock.
\newblock {\em Optica}, 6(5):680--685, 2019.

\bibitem{he2019self}
Yang He, Qi-Fan Yang, Jingwei Ling, Rui Luo, Hanxiao Liang, Mingxiao Li, Boqiang Shen, Heming Wang, {Kerr}y Vahala, and Qiang Lin.
\newblock Self-starting bi-chromatic {LiNbO$_3$} soliton microcomb.
\newblock {\em Optica}, 6(9):1138--1144, 2019.

\bibitem{lv2025broadband}
Xiaomin Lv, Binbin Nie, Chen Yang, Rui Ma, Ze~Wang, Yanwu Liu, Xing Jin, Kaixuan Zhu, Zhenyu Chen, Du~Qian, et~al.
\newblock Broadband microwave-rate dark pulse microcombs in dissipation-engineered {LiNbO$_3$} microresonators.
\newblock {\em Nat. Commun.}, 16(1):2389, 2025.

\bibitem{nie2025soliton}
Binbin Nie, Xiaomin Lv, Chen Yang, Rui Ma, Kaixuan Zhu, Ze~Wang, Yanwu Liu, Zhenyu Xie, Xing Jin, Guanyu Zhang, et~al.
\newblock Soliton microcombs in x-cut {LiNbO$_3$} microresonators.
\newblock {\em eLight}, 5(1):15, 2025.

\bibitem{song2024octave}
Yunxiang Song, Yaowen Hu, Xinrui Zhu, Kiyoul Yang, and Marko Lon{\v{c}}ar.
\newblock Octave-spanning {Kerr} soliton frequency combs in dispersion-and dissipation-engineered lithium niobate microresonators.
\newblock {\em Light: Sci. Appl.}, 13(1):225, 2024.

\bibitem{shen2020integrated}
Boqiang Shen, Lin Chang, Junqiu Liu, Heming Wang, Qi-Fan Yang, Chao Xiang, Rui~Ning Wang, Jijun He, Tianyi Liu, Weiqiang Xie, et~al.
\newblock Integrated turnkey soliton microcombs.
\newblock {\em Nature}, 582(7812):365--369, 2020.

\bibitem{wang2025reconfigurable}
Yufei Wang, Kun Liao, Kuo Zhang, Zhuochen Du, Ze~Wang, Bo~Ni, Tianyu Xu, Shuai Feng, Yan Yang, Qi-Fan Yang, et~al.
\newblock Reconfigurable versatile integrated photonic computing chip.
\newblock {\em eLight}, 5(1):20, 2025.

\bibitem{xu2021mitigating}
Yuntao Xu, Mohan Shen, Juanjuan Lu, Joshua~B Surya, Ayed~Al Sayem, and Hong~X Tang.
\newblock Mitigating photorefractive effect in thin-film lithium niobate microring resonators.
\newblock {\em Opt. Express}, 29(4):5497--5504, 2021.

\bibitem{kippenberg2004kerr}
TJ~Kippenberg, SM~Spillane, and KJ~Vahala.
\newblock {Kerr}-nonlinearity optical parametric oscillation in an ultrahigh-q toroid microcavity.
\newblock {\em Phys. Rev. Lett.}, 93(8):083904, 2004.

\bibitem{RN3784}
Chen Yang, Shuo Yang, Fan Du, Xianhong Zeng, Beichen Wang, Zijiao Yang, Qiang Luo, Rui Ma, Ru~Zhang, Di~Jia, Zhenzhong Hao, Yongnan Li, Qifan Yang, Xu~Yi, Fang Bo, Yongfa Kong, Guoquan Zhang, and Jingjun Xu.
\newblock 1550-nm band soliton microcombs in ytterbium-doped lithium-niobate microrings.
\newblock {\em Laser Photonics Rev.}, 17(9):2200510, 2023.

\bibitem{RN4288}
Jiangang Zhu, Şahin~Kaya Özdemir, Lina He, and Lan Yang.
\newblock Controlled manipulation of mode splitting in an optical microcavity by two rayleigh scatterers.
\newblock {\em Opt. Express}, 18(23):23535--23543, 2010.

\bibitem{jeong2020ultralow}
Dongin Jeong, Dohyeon Kwon, Igju Jeon, In~Hwan Do, Jungwon Kim, and Hansuek Lee.
\newblock Ultralow jitter silica microcomb.
\newblock {\em Optica}, 7(9):1108--1111, 2020.

\bibitem{lao2023quantum}
Chenghao Lao, Xing Jin, Lin Chang, Heming Wang, Zhe Lv, Weiqiang Xie, Haowen Shu, Xingjun Wang, John~E Bowers, and Qi-Fan Yang.
\newblock Quantum decoherence of dark pulses in optical microresonators.
\newblock {\em Nat. Commun.}, 14(1):1802, 2023.

\bibitem{wang2025compact}
Yuanlei Wang, Ze~Wang, Chenghao Lao, Tianyu Xu, Yinke Cheng, Zhenyu Xie, Junqi Wang, Haoyang Luo, Xin Zhou, Bo~Ni, et~al.
\newblock Compact turnkey soliton microcombs at microwave rates via wafer-scale fabrication.
\newblock {\em arXiv preprint arXiv:2502.10941}, 2025.

\bibitem{RN4284}
Rui Bao, Zhiwei Fang, Jian Liu, Zhaoxiang Liu, Jinming Chen, Min Wang, Rongbo Wu, Haisu Zhang, and Ya~Cheng.
\newblock An erbium-doped waveguide amplifier on thin film lithium niobate with an output power exceeding 100 mw.
\newblock {\em Laser Photonics Rev.}, 19(1):2400765, 2025.

\bibitem{RN4283}
Yunxiang Song, Xinrui Zhu, Xiangying Zuo, Guanhao Huang, and Marko Lončar.
\newblock Stable gigahertz- and mmwave-repetition-rate soliton microcombs on x-cut lithium niobate.
\newblock {\em Optica}, 12(5):693--701, 2025.

\bibitem{xue2015mode}
Xiaoxiao Xue, Yi~Xuan, Yang Liu, Pei-Hsun Wang, Steven Chen, Jian Wang, Dan~E Leaird, Minghao Qi, and Andrew~M Weiner.
\newblock Mode-locked dark pulse {Kerr} combs in normal-dispersion microresonators.
\newblock {\em Nat. Photonics}, 9(9):594--600, 2015.

\bibitem{zhu2025ultra}
Kaixuan Zhu, Xinrui Luo, Yuanlei Wang, Ze~Wang, Tianyu Xu, Du~Qian, Yinke Cheng, Junqi Wang, Haoyang Luo, Yanwu Liu, et~al.
\newblock Ultra-broadband soliton microcombs in resonantly-coupled microresonators.
\newblock {\em arXiv preprint arXiv:2503.02022}, 2025.

\bibitem{shu2022microcomb}
Haowen Shu, Lin Chang, Yuansheng Tao, Bitao Shen, Weiqiang Xie, Ming Jin, Andrew Netherton, Zihan Tao, Xuguang Zhang, Ruixuan Chen, et~al.
\newblock Microcomb-driven silicon photonic systems.
\newblock {\em Nature}, 605(7910):457--463, 2022.

\bibitem{corcoran2025optical}
Bill Corcoran, Arnan Mitchell, Roberto Morandotti, Leif~K Oxenl{\o}we, and David~J Moss.
\newblock Optical microcombs for ultrahigh-bandwidth communications.
\newblock {\em Nat. Photonics}, 19(5):451--462, 2025.

\bibitem{RN4281}
Yunxiang Song, Yaowen Hu, Xinrui Zhu, Keith Powell, Letícia Magalhães, Fan Ye, Hana~K. Warner, Shengyuan Lu, Xudong Li, Dylan Renaud, Norman Lippok, Di~Zhu, Benjamin Vakoc, Mian Zhang, Neil Sinclair, and Marko Lončar.
\newblock Integrated electro-optic digital-to-analogue link for efficient computing and arbitrary waveform generation.
\newblock {\em Nat. Photonics}, 2025.

\end{thebibliography}

\end{document}